\documentclass{aa}  

\usepackage{graphicx}
\usepackage{txfonts}
\usepackage{caption}
\usepackage{xcolor}
\usepackage{hyperref}
\usepackage{comment}
\makeatletter
\renewcommand*\aa@pageof{, page \thepage{} of \pageref*{LastPage}}
\makeatother
\begin{document} 

   \title{High angular resolution near-ultraviolet polarization imaging\\ of the Herbig Ae/Be star LK-H$\alpha$-233}

   \author{F. Marin\inst{1}\thanks{\href{mailto:frederic.marin@astro.unistra.fr}{frederic.marin@astro.unistra.fr}}
    }
    \institute{Universit\'e de Strasbourg, CNRS, Observatoire astronomique de Strasbourg, UMR 7550, F-67000 Strasbourg, France
    }

   \date{Received March 24, 2025; accepted May 20, 2025}

 
  \abstract
   {Herbig Ae/Be stars are young, pre-main-sequence stars that provide critical insights into the processes of stellar formation, early stellar evolution and protoplanetary disks.}
   {Two of the key features of such stars are their circumstellar dusty disk and bipolar ionized outflows, which are key components for understanding planet formation processes and energy/matter deposition in the interstellar medium, respectively. In this context, imaging polarimetry is probably the sharpest tool to characterize the various structures and dynamics around the central star, due to the sensitivity of polarization to the morphology of the emitting, scattering and absorbing media.}
   {We take advantage of never published, near-ultraviolet polarimetric data of LK-H$\alpha$-233 taken by the Faint Object Camera aboard the Hubble Space Telescope in 1991, 1994 and 1995, which remained dormant in the archives despite their quality. Using the most recent and robust reduction pipeline for this instrument, we obtained high spatial resolution (0.0287 $\times$ 0.0287~arcsecond$^2$) maps of this object at 4118~\AA, together with polarimetric measurements.}
   {A dark lane, bisecting the approaching and receding polar outflows, suggests the presence of a circumstellar disk or dust torus, obscuring the pre-main sequence star and collimating the ejecta. Polarization reveals that the outflows have an X-shape structure with a significant centro-symmetric pattern in polarization angle, indicating that the outflows are both hollow and scattering the emission from the buried star. We constrain the half-opening angle of both the outflows and circumstellar disk, determine the inclination of the system and estimate the obscured star's intrinsic flux}
   {This study highlights the importance of high-resolution polarimetric observations in understanding the complex environment around Herbig Ae/Be stars and advocates for future similar instruments.}


   \maketitle
%

\section{Introduction}

The Faint Object Camera (FOC) aboard the Hubble Space Telescope (HST) was a pivotal instrument in astronomical research, known for its high angular resolution imaging and polarimetric capabilities \citep{Macchetto1982,Nota1996}. Operating in the 1150 to 6500~\AA\, wavelength range, the FOC utilized a combination of advanced optics and detectors to achieve unprecedented spatial resolution (14 $\times$ 14 ~arcseconds$^2$ field-of-view with pixel dimensions as small as 0.01435 $\times$ 0.01435~arcseconds$^2$), crucial for resolving fine structures in distant astronomical objects. Its ability to perform polarimetry allowed to measure the polarization of light, providing insights into the scattering mechanisms and magnetic fields within various astrophysical environments. The high angular resolution of the FOC, combined with its polarimetric functionality, made it an exceptionally powerful diagnostic tool for studying complex systems such as gravitational lenses \citep{Nguyen1999}, active galactic nuclei (AGNs, \citealt{Capetti1995,Thomson1995,Kishimoto2002}), supernovae \citep{Sparks1999} and other intricate cosmic phenomena. These capabilities enabled detailed investigations into the morphology and dynamics of celestial objects, until it was removed during the fourth servicing mission (SM3B) in March 2002. 

Among the sources that were observed by the HST/FOC, AGNs are the most numerous, followed by supernovae and by a few more singular objects. Among them is LK-H$\alpha$-233. This source was part of a calibration proposal, so it is not included in the catalog of sources observed for scientific purposes. Its observation was thus forgotten and never analyzed in a scientific context. LK-H$\alpha$-233 is not the only source to have been consigned to oblivion in the archives, as illustrated by the project undertook by \citet{Barnouin2023}. However, the work of these authors focuses on AGNs. In this paper, we focus on the only Herbig Ae/Be star ever observed by the FOC.

Herbig Ae/Be stars are a subclass of pre-main-sequence stars that occupy a crucial phase in the evolutionary timeline of stellar development. These young, intermediate-mass stars, with masses ranging between 1.5 and 10 solar masses \citep{Vioque2018,Brittain2023}, are characterized by their strong emission lines \citep{Manoj2006}, infrared excesses \citep{Hartmann1993}, and association with reflection nebulae \citep{Herbig1960,Neckel1987}. Discovered and classified by George Herbig in the 1960s \citep{Herbig1960}, these stars serve as prime laboratories for studying the processes of star formation and early stellar evolution.

The defining features of Herbig Ae/Be stars include their surrounding protoplanetary disks and the presence of bipolar jets and outflows \citep{Hillenbrand1992}. These circumstellar disks, composed of gas and dust, are the birthplaces of planetary systems, making Herbig Ae/Be stars critical for understanding planet formation mechanisms \citep{Brittain2023}. The interaction between the star and its disk leads to complex phenomena such as accretion, disk winds, and the formation of circumstellar structures, all of which can be studied through multi-wavelength observations. For examples, the Near-Infrared Camera and Multi Object Spectrometer (NICMOS) on the HST provided the first resolved scattered light images of dusty disks around those pre-main sequence stars \citep{Weinberger1999} ; the Atacama Large Millimeter/submillimeter Array (ALMA) can spatially resolve the disk emission in several Herbig Ae/Be stars thanks to a big step forward in terms of spatial resolution (about 0.05 -- 0.1~arcseconds, \citealt{Andrews2018}) ; the Gaia satellite, operating in the optical band, made it possible to homogeneously determine the masses, ages, luminosities, distances, photometric variabilities and infrared excesses from large samples of Herbig Ae/Be stars \citep{Vioque2018}, and so on. Polarimetry has also been used effectively to study those stars, from determining the dust composition in the disks \citep{Murakawa2010b} to constraining the accretion processes and bridging them to T Tauri stars, the lower mass counterparts to Herbig Ae/Be stars \citep{Vink2015,Ababakr2017}.

LK-H$\alpha$-233, the star we are interested in here, has been included in many of those studies. Situated at a distance of 880~pc \citep{Testi1998,Manoj2006}, where 1~arcsec = 0.0042664~pc = 0.0139151~ly = 880~au, it always was a good target for imaging and polarimetry. \citet{Aspin1985}, using imaging polarimetry and photometry in three broad bands (B, V, R) discovered that the central star is hidden behind a large obscuring region and that the associated faint nebulosity seen in optical \citep{Herbig1960} is in fact a bipolar outflow that is highly polarized, suggesting scattering from an optically-thin medium. This discovery was confirmed by near-infrared speckle interferometry \citep{Leinert1993} and HST's Space Telescope Imaging Spectrograph (STIS) and Wide Field Planetary Camera (WFPC2) observations \citep{Melnikov2008}. Those authors also reported the presence of a dark lane caused either by a circumstellar disk or a dust torus that is extincting the star itself, and helped to characterize the bipolar outflows. 

In this paper, we explore what the forgotten near-ultraviolet polarimetric data of the FOC can tell us about the circumstellar environment of LK-H$\alpha$-233. In Sect.~\ref{Archives}, we describe how the data were retrieved and reduced, and present the three different observations made with the FOC. In Sect.~\ref{Analysis}, we analyze in great details the spatially resolved polarization of LK-H$\alpha$-233 and derive constraints on the outflows and cirumstellar disk geometry. We discuss our results in Sect.~\ref{Discussion} and make a link between the general geometry of the Herbig Ae/Be stars and AGNs. We conclude our paper in Sect.~\ref{Conclusion}, highlighting the need for future high angular resolution polarimeters to study protoplanetary disks and stellar evolution.

\section{The forgotten archives}
\label{Archives}

Originally, the principal investigators of the forgotten archives, Francesco Paresce and Philip  Hodge, wanted to determine the FOC unpolarized transmission coefficients in orbit to a relative accuracy of 0.3\% by observing a standard extended unpolarized ($<$ 0.15\%) source (NGC~5272), as well as the effective polarimeter position angle offset and polarized transmission coefficients to a relative accuracy of 3\% by observing a standard polarized source (LK-H$\alpha$-233) through the blue polarizers. The LK-H$\alpha$-233 observations were analyzed in an instrumental context, but never in a scientific context, and they have remained dormant in the archives ever since. We also verified that no other similar objects had also been forgotten and we confirm that this is the only unpublished, never-analyzed observation of a pre-main sequence star with the HST/FOC.

\subsection{Data reduction}
\label{Archives:pipeline}

Three LK-H$\alpha$-233 observations were conducted using the FOC aboard HST : one in 1991, another one in 1994 and the last one in 1995. For all observations, the f/96 mode was used ($512 \times 512$ pixel format) without zoom, offering a spatial resolution of $0.01435 \times 0.01435$ arcseconds$^2$ per pixel, equivalent to an effective resolution of 6.116 $\times$ 10$^{-5}$~pc ($\approx$ 12.6~AU) for a $7 \times 7$ arcseconds field of view. However, the final effective spatial resolution will be limited by binning and smoothing, and thus slightly degraded (see next paragraph). Exposures were taken with the F430W broadband blue filter centered on 3960~\AA\, (bandpass 870~\AA), along with three polarizing Rochon prisms (POL0, POL60, POL120). In contrary to, e.g. the HST/FOC observation of M87 \citep{Marin2024b}, no FOC warming up, preliminary shots to precisely place the source on the same detector region, nor internal flats to remove reseau marks which were used to calibrate geometrical distortion of the cathode ray tube, were taken. This will be important in the analysis of the results.

To reduce the data downloaded from the MAST HST Legacy Archive\footnote{\url{https://archive.stsci.edu/missions-and-data/hst}}, we followed the guidelines and used the automated, generalized reduction pipeline presented in \citet{Barnouin2023}. We refer the reader to this paper and the other of this series (\citealt{Barnouin2024}, \citealt{Marin2024b}) for technical details. Here, we only describe the parametrization used to obtain the fully reduced polarization maps. After downloading the data, the raw POL0, POL60, and POL120 images were cropped to show the same region of interest. Background estimation was done by analyzing the intensity histogram of each image using the Freedman-Diaconis sampling rule. A Gaussian fit was applied to the histogram, with the mean value representing the background level. Images were aligned to a precision of approximately 0.1 pixels, then binned at the Nyquist frequency, i.e. 2 $\times$ 2 pixels. As a reminder, the Nyquist–Shannon theorem states that an analog signal can be digitized without aliasing error if and only if the sampling rate is greater than or equal to twice the highest frequency component in a given signal \citep{Shannon1949}. This allows us to maximize the spatial resolution of the map whilst retaining trustable spatial (and polarization) information. The images were then smoothed using a Gaussian kernel with a full-width at half-maximum of 2 pixels, twice the size of a resampled pixel. For each pixel, the Stokes parameters I, Q, and U (with their associated uncertainties) were computed, along with the debiased polarization degree $P$ and the electric vector position angle $\theta$. The final maps were rotated to have North up ($\theta = 0^\circ$), with the rotation of $\theta$ following the IAU convention (the value for the electric-vector position angle of polarization starts from North and increases through East). No deconvolution was applied during the process. The resulting images have a spatial resolution of 0.0287~arcseconds ($\approx$ 25.2~AU), which make them the highest angular resolution polarimetric maps of this object to date, almost ten times more resolved than the total flux HST/STIS and HST/WFPC2 observations presented in \citet{Melnikov2008}.

However, as we will see in the following subsections, this high spatial resolution is very demanding in photons count to obtain statistically credible polarization measurements in each individual pixel. More polarimetric information can be obtained using maps with a slightly lower spatial resolution to maximize the counts in each pixel and thus detect with more certainty the polarization of LK-H$\alpha$-233. Therefore, the high resolution images to be showed in the following will be accompanied by medium resolution images, where a spatial binning of 0.1 arcseconds has been performed, with a Gaussian smoothing of 0.2 arcseconds. All other reduction details are the same between the high and medium resolution final images. Full scale images are presented in Appendix for better visibility.

\subsection{1991's observation}
\label{Archives:1991}

The first observation (Program ID: 3873) was taken between November 1991, 20 and November 1991, 22. Each polarizer accumulated approximately 3.0 ks, resulting in a total exposure time of about 2.5 hours in polarimetric mode. This observation has the distinction of having been acquired when the HST still had a spherical aberration on its primary mirror. This flaw caused the mirror to be too flat by about 2.2~$\mu$m at its edges, leading to blurred and distorted images \citep{White1991}. 

\begin{figure*}
\centering
\includegraphics[width=0.48\textwidth]{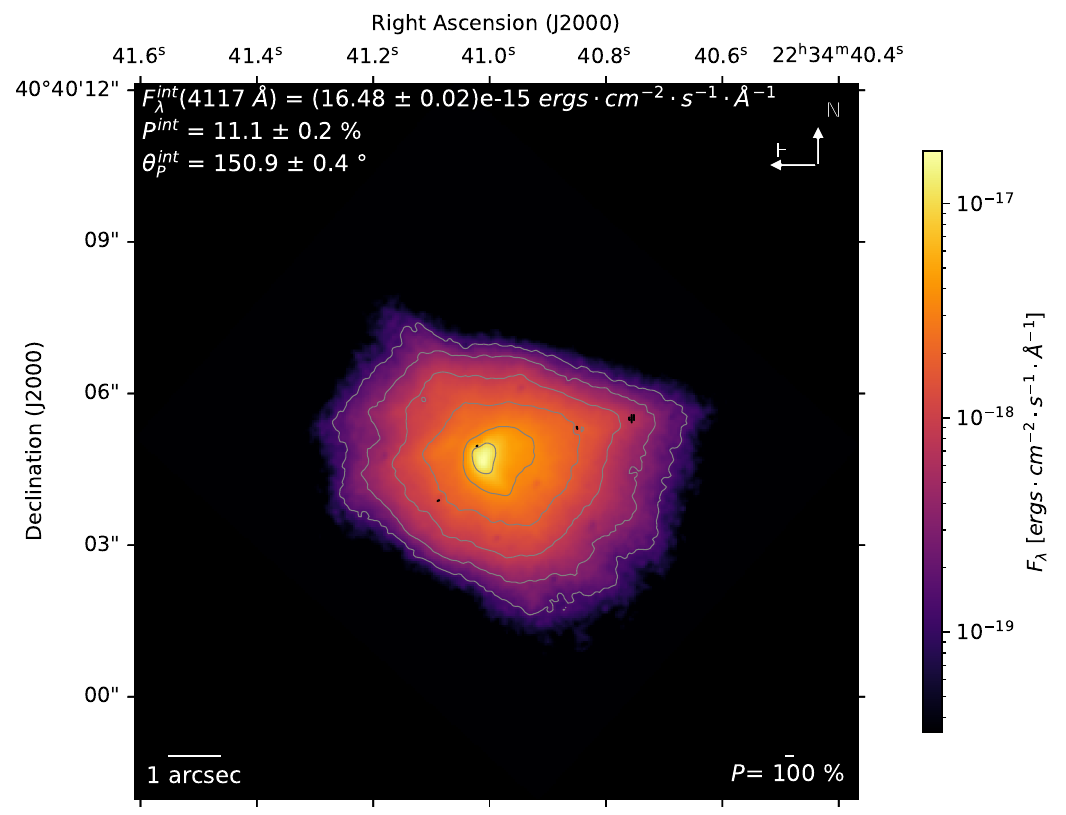}
\includegraphics[width=0.48\textwidth]{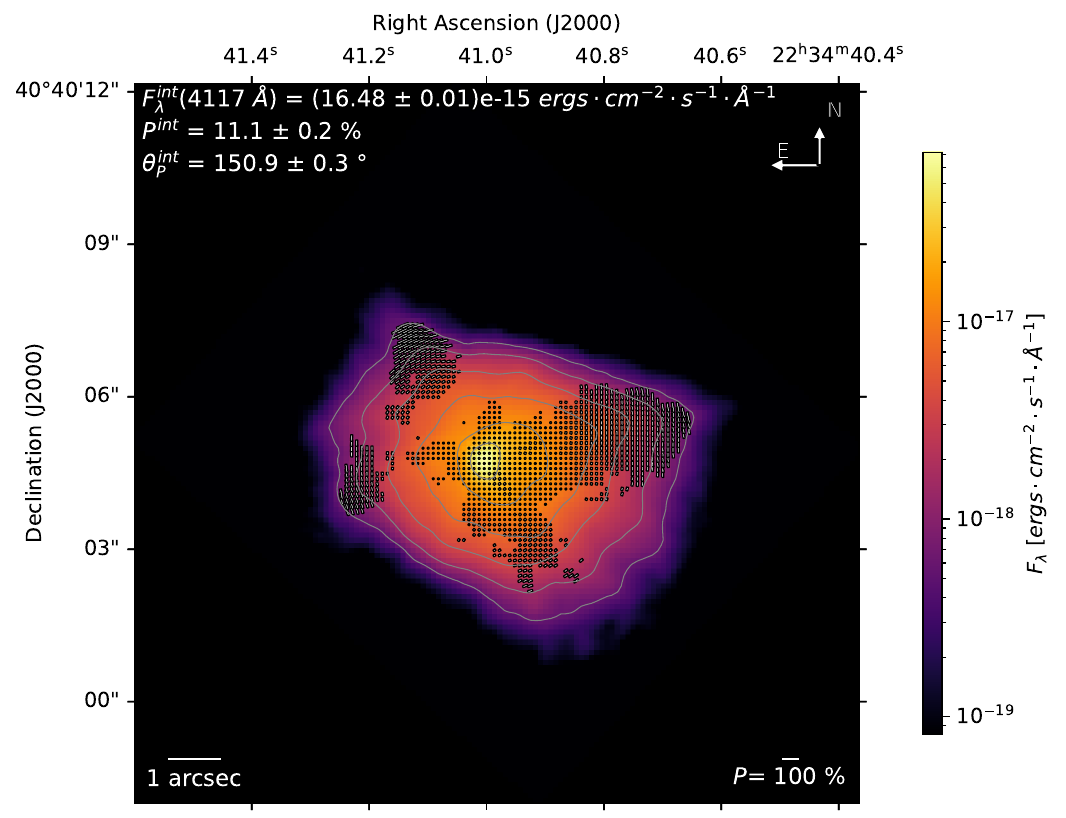}
\caption{Left : 1991's pre-COSTAR HST/FOC observation of LK-H$\alpha$-233 resampled according to the Nyquist–Shannon sampling theorem, i.e. 2 $\times$ 2 pixels$^2$ (0.0287 $\times$ 0.0287~arcseconds$^2$), in order to have individual pixels with meaningfully polarization measurements. The total flux is color-coded and expressed in units of erg cm$^{-2}$ s$^{-1}$ \AA$^{-1}$. Polarization vectors are displayed for $\left[\text{S/N}\right]_P \geq 3$, even if at this spatial resolution they are difficult to resolve by eyes (larger images are presented in appendices). Total flux contours are displayed for 0.8\%, 2\%, 5\%, 10\%, 20\% and 50\% of the maximum flux. On the top-left corner, we indicate the total flux $F$, polarization degree $P$ and polarization angles $\theta$ values, integrated over the whole field of view (7 $\times$ 7~arcseconds$^2$ in this image). Following the IAU convention, the North is up and the polarization angle positively rotates towards the East. Right : same image but with a spatial binning of 0.1 arcsecond per pixel. The polarization pattern highlighted by the vectors is then much easier to detect.}
\label{Fig:1991_I}%
\end{figure*}

Despite this defect, we present the 1991's observed polarization maps of LK-H$\alpha$-233 in Fig.~\ref{Fig:1991_I}. The maps show the color-coded total flux spatially sampled at the Nyquist frequency (left) and at 0.1 arcsecond per pixel (right). The polarization information is overplotted on the flux using vectors whose length is proportional to $P$ and whose orientation traces $\theta$. Due to the large spatial resolution of the map, the vectors are barely visible, but full scale images are presented in appendix. We see that LK-H$\alpha$-233 has an almost isotropic flux distribution, with a slight elongation in the North-East/South-West direction (position angle $\approx$ 66$^\circ$, in agreement with the 60 - 70$^\circ$ value found by \citealt{Corcoran1998}). More than 50\% to the observed flux comes from the inner, sub-arcsecond region. A careful observer will notice a network of gray dots in the source flux: these are the instrument's reseau markings, which could not be corrected due to the absence of prior internal flats. These points are often associated with polarization vectors which are only observational artifacts. No polarization vector intrinsic to the source is reported in the high resolution image but a centro-symmetric polarization pattern is clearly visible in the medium resolution map. Such polarization pattern will be analyzed in details in Sect.~\ref{Analysis}.

The source presents an integrated ($7 \times 7$~arcseconds$^2$) total flux of $\sim$ 1.6 $\times$ 10$^{-14}$ erg cm$^{-2}$ s$^{-1}$ \AA$^{-1}$. It is difficult to assess the validity of this flux measurement as no continuum values are reported in the literature at this waveband, nor in the optical. The integrated polarization, 11.1\% $\pm$ 0.2\% at 150.9$^\circ$ $\pm$ 0.4$^\circ$, is, however, entirely consistent with the values reported by \citet{Vrba1979}, i.e., 10.8\% $\pm$ 0.1\% at 153.9$^\circ$ $\pm$ 0.2$^\circ$ in the B-band with a 10~arcsecond aperture. We note that this polarization angle is perpendicular to the position angle of the outflows, likely indicating perpendicular scattering from stellar photons onto the ejecta. It thus appears that the HST aberration, despite blurring the details of the observed source, does not distort the measured polarization values too much (as it was proven by the observation of the asctive galactic nucleus NGC~1068 before and after correction of the aberration, see \citealt{Capetti1995}).

\subsection{1994's observation}
\label{Archives:1994}

The second observation (Program ID: 5522) was taken on November 1994, 1. Each polarizer accumulated approximately 1.5 ks, resulting in a total exposure time of about 1.2 hours in polarimetric mode. This observation, and the one obtained in 1995, benefits from the installation of the Corrective Optics Space Telescope Axial Replacement (COSTAR) in 1993, during a servicing mission. The COSTAR was designed to compensate for the primary mirror's flaw by deploying small, precisely shaped mirrors into the optical paths of several of Hubble's scientific instruments, such as the FOC. These corrective mirrors redirected and refocused light to eliminate the spherical aberration \citep{Hartig1993}.

\begin{figure*}
\centering
\includegraphics[width=1\textwidth]{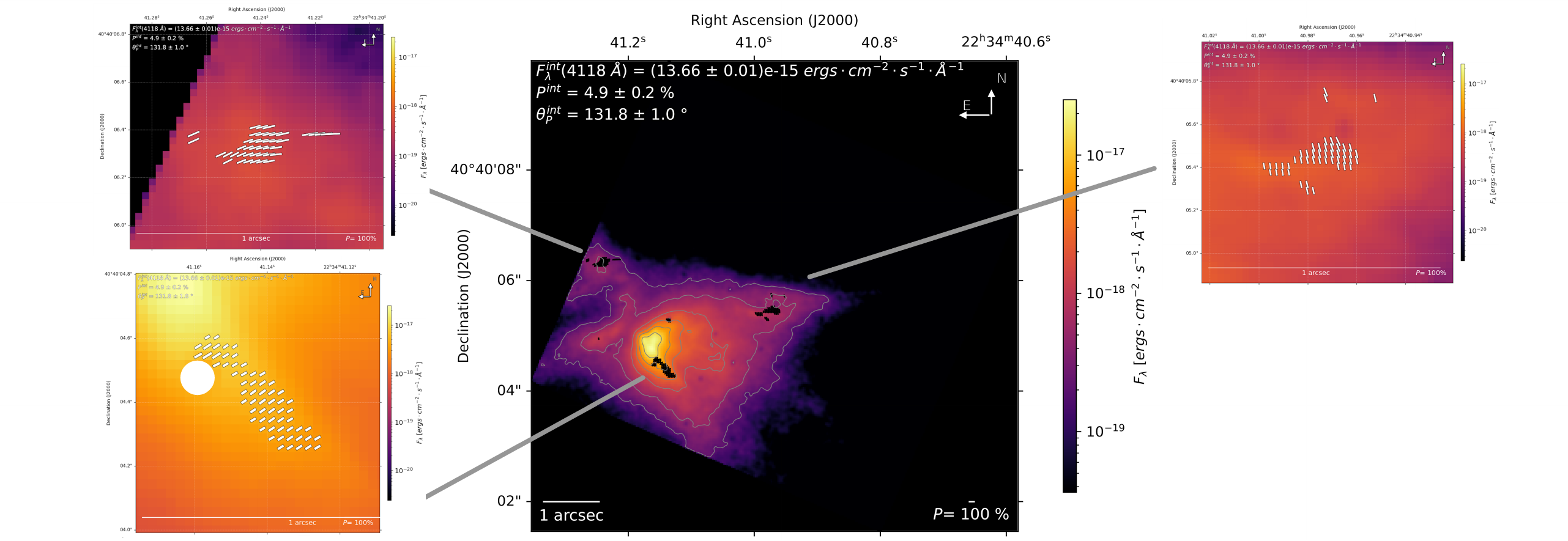}
\includegraphics[width=0.5\textwidth]{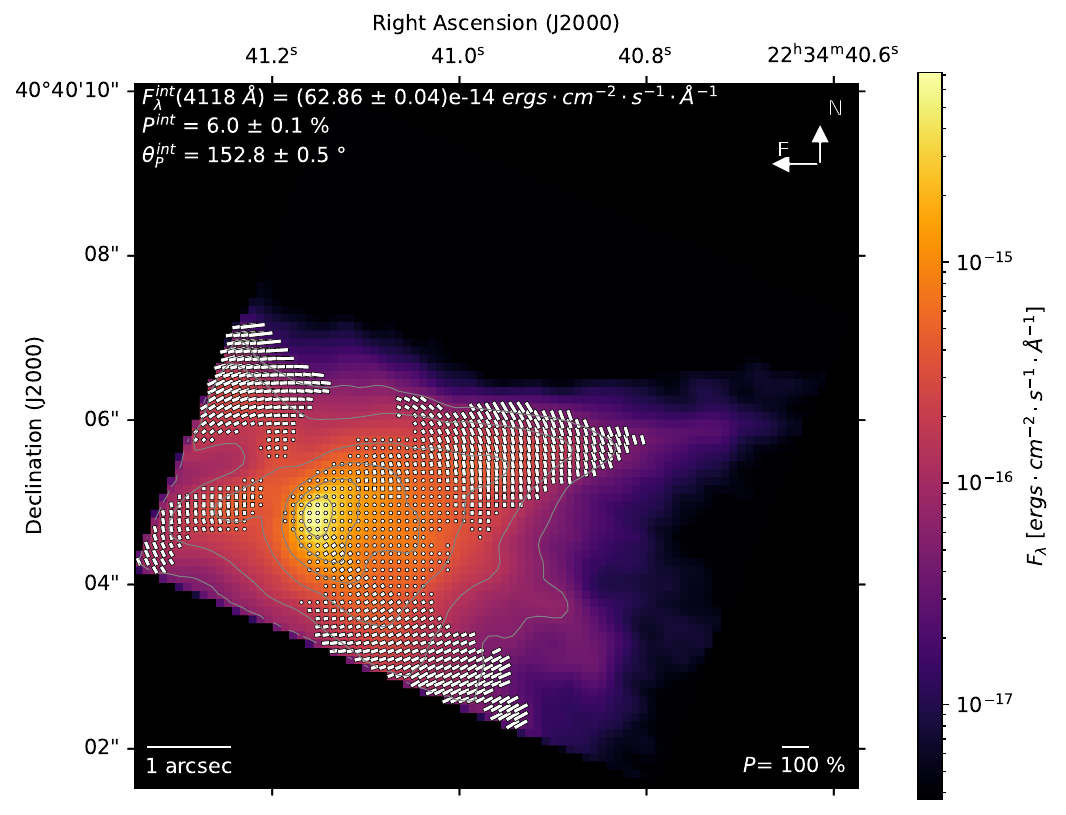}
\caption{Same as Fig.~\ref{Fig:1991_I} but for the 1994's post-COSTAR HST/FOC observation of LK-H$\alpha$-233. The top image is the high resolution map, with inserts to show where $\left[\text{S/N}\right]_P \geq 3$ polarization vectors are detected. The white dot corresponds to a reseau mark that was masked in the reduction process. The bottom image is the medium resolution map. Due to slightly incorrect pointing, the source appears in the bottom-left corner of the FOC field-of-view and is partly cut off. The huge increase in image sharpness due to the COSTAR corrective optics package is striking.}
\label{Fig:1994_I}%
\end{figure*}

The 1994's observation of LK-H$\alpha$-233 is presented in Fig.~\ref{Fig:1994_I}. The increase in sharpness is blatant. The source is now resolved into different components, even though the pointing was a bit erroneous. The source appears on the bottom-left corner of the FOC field-of-view, and it was unfortunately partially cut. Nevertheless, the central and western components of the source, the star and the approaching, blueshifted outflow (see \citealt{Melnikov2008}), respectively, are clearly visible. The position angle of the outflows is confirmed at 66$^\circ$ $\pm$ 1$^\circ$. Now, a rectilinear zone of weaker flux clearly separates the eastern and western winds, which spread in a conical fashion from their apex (the unresolved pre-main sequence star). We also observe several polarization vectors that are distinctively different from the reseau marks, indicating that $\left[\text{S/N}\right]_P \geq 3$ polarized pixels are detected. The medium resolution image also shows the centro-symmetric polarization pattern, but with a total number of pixels where polarization is clearly measured higher than in the distorted 1991 image. The flux, polarization degree, and polarization angle reported on the medium resolution image appear to be a bit different from that of the high resolution map. This effect is due to the binning and smoothing applied to the Stokes $Q$, $U$, and $I$ parameters prior to computing the polarization. This processing step alters the local vector structure (particularly in regions with steep polarization gradients or low signal) by averaging the Stokes parameters over larger areas, which can lead to a reduction in polarization degree and shifts in polarization angle, even though the underlying physical signal remains the same. The same will apply to the 1995 observation but not to the 1991 pointing, as the lack of COSTAR corrections already blurred the image.

Taken from the high resolution image, where the smoothing is very light and therefore only minimally affects regions near the edge of the field of view, the integrated flux is $\sim$ 1.4 $\times$ 10$^{-14}$ erg cm$^{-2}$ s$^{-1}$ \AA$^{-1}$. It is marginally lower than in 1991, but the fact that a fraction of the source has been cut-off must be accounted for here. Flux variation is not an unusual behavior for Herbig Ae/be stars, as such objects are known to vary both in continuum and line fluxes \citep{Perez1993,Mendigutia2011}. We also note that the observed integrated polarization is 4.9\% $\pm$ 0.2\% at 131.8$^\circ$ $\pm$ 1.0$^\circ$. The polarization degree and angle are very different from the integrated value measured in 1991 (although they remain perpendicular to the outflows), which is also something that can be expected from an Herbig Ae/be star. Indeed, \citet{Jain1995} have demonstrated that polarimetric variability is a common feature of this class of objects, with variability spanning from months to years. However, caution must be taken with this apparent variability of polarization since part of the source extends beyond the field-of-view of the instrument, something that can alter the integrated polarization measurement.

\subsection{1995's observation}
\label{Archives:1995}

The third and last observation (Program ID: 5522, same as the second observation) was taken between June 1995, 17 and June 1995, 18. Each polarizer accumulated approximately 1.5 ks, resulting in a total exposure time of about 1.2 hours in polarimetric mode.

\begin{figure*}
\centering
\includegraphics[width=1\textwidth]{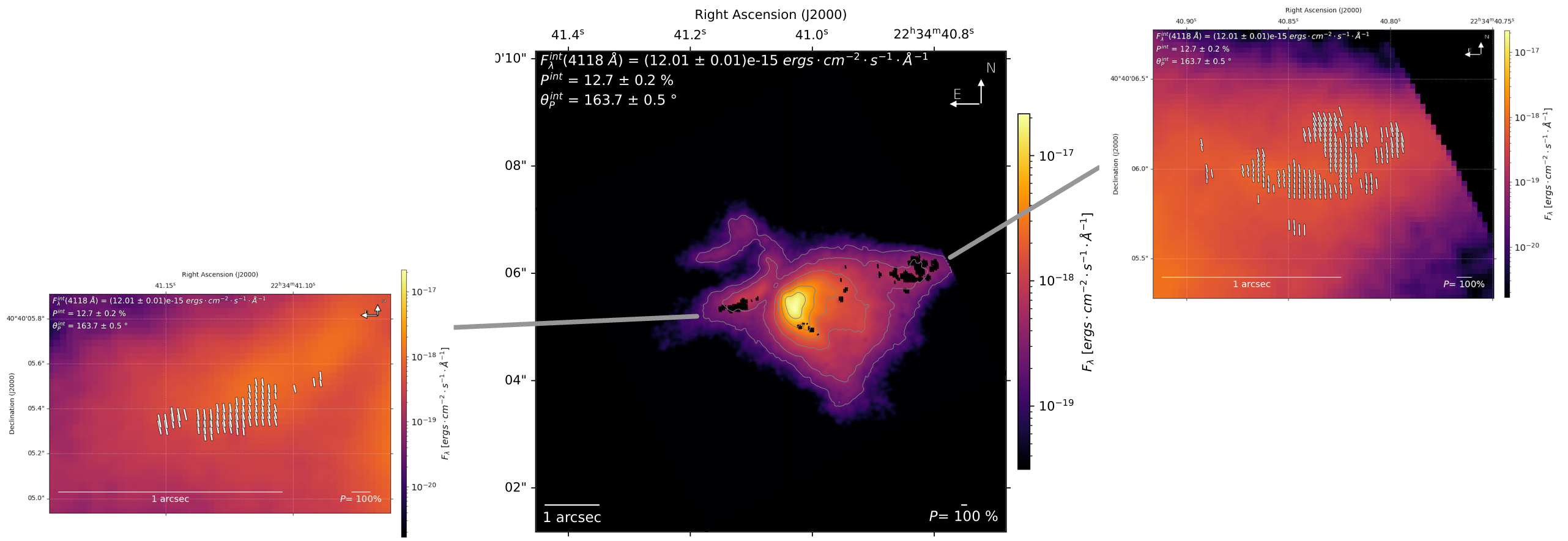}
\includegraphics[width=0.5\textwidth]{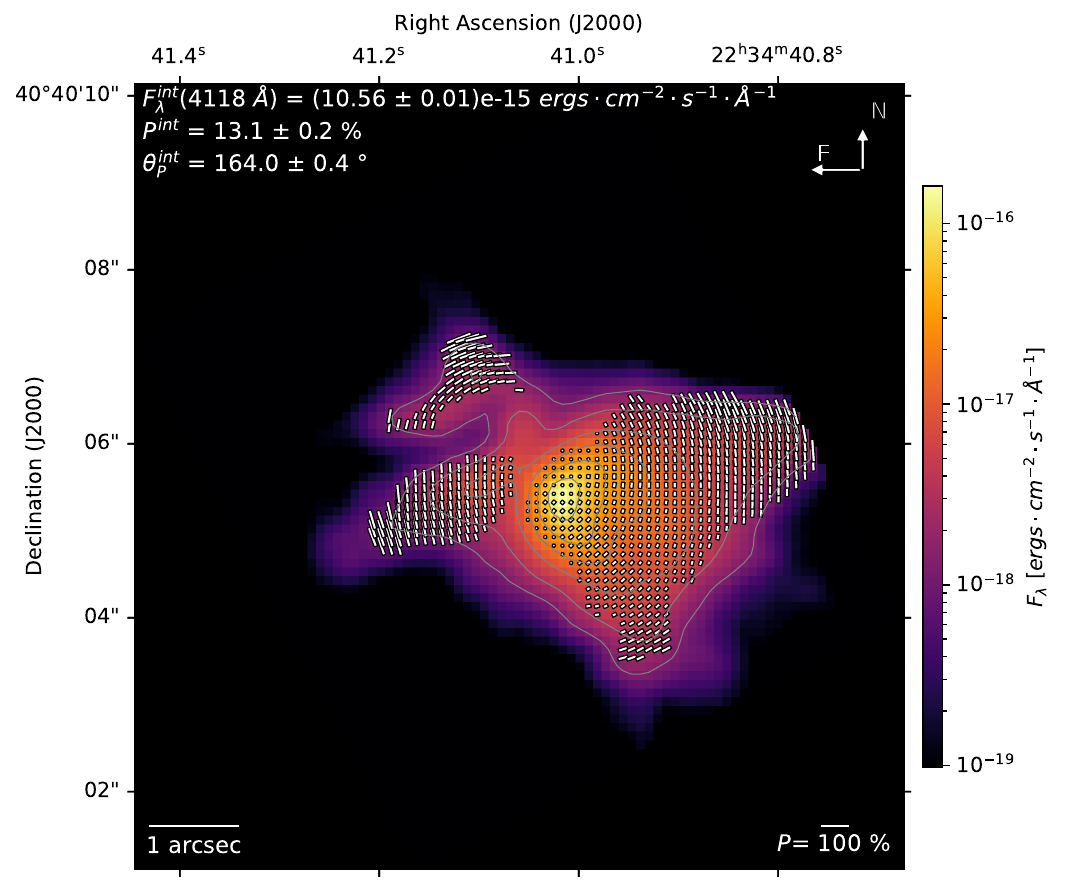}
\caption{Same as Fig.~\ref{Fig:1994_I} but for the 1995's post-COSTAR HST/FOC observation of LK-H$\alpha$-233. Due to slightly incorrect pointing, the western extremity of the source is slightly cut off the field-of-view.}
\label{Fig:1995_I}%
\end{figure*}

The 1995's observation of LK-H$\alpha$-233 is presented in Fig.~\ref{Fig:1995_I}. The pointing accuracy was better but, again, a fraction of the source is cut. This time, it is only the extremity of the western wind that is cut, something that should not hinder the analysis, unlike the 1994 observation where a much larger fraction of the object was out of the field-of-view. The source presents similar flux geometry (and intensity), as highlighted by the flux contours on Figs.~\ref{Fig:1994_I} and \ref{Fig:1995_I}.

The integrated flux is $\sim$ 1.2 $\times$ 10$^{-14}$ erg cm$^{-2}$ s$^{-1}$ \AA$^{-1}$, slightly lower than the 1994 observation. The observed integrated polarization is 12.7\% $\pm$ 0.2\% at 163.7$^\circ$ $\pm$ 0.5$^\circ$. The polarization is, again, different from the 1991 and 1994 values, but the polarization angle is still perpendicular to the ejecta. Over three epochs, the polarization shows clear signs of variability, but it is less certain for the total flux .

\section{A deeper analysis of the data}
\label{Analysis}

\subsection{A polarized view}
\label{Analysis:pol_map}

\begin{figure}
\centering
\includegraphics[width=1\columnwidth]{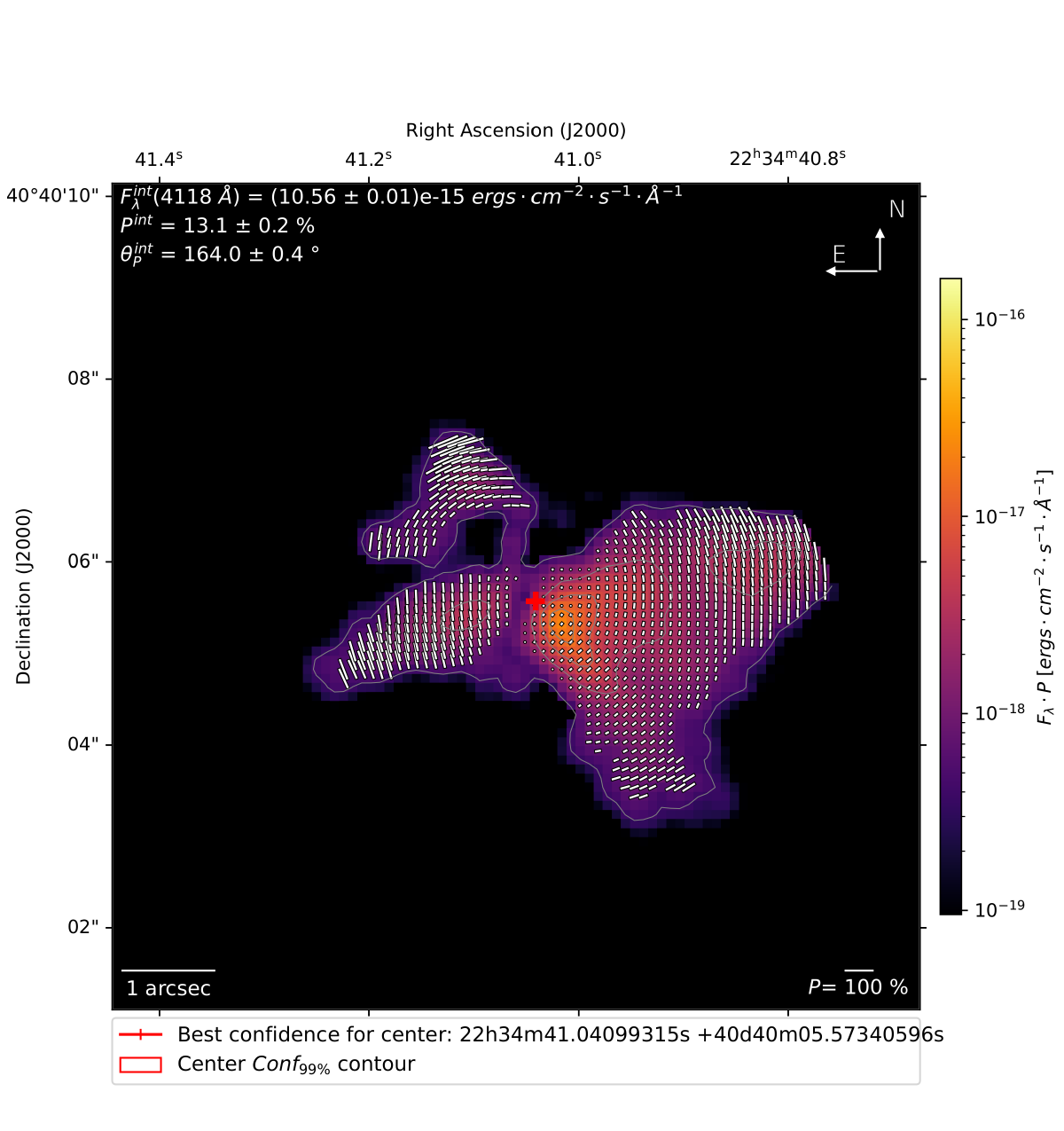}
\caption{1995's medium resolution post-COSTAR HST/FOC observation of LK-H$\alpha$-233, seen in polarized flux. The red cross shows the estimated location of the hidden star.}
\label{Fig:Center}%
\end{figure}

We start by examining the 1995's HST/FOC image in polarized flux (the multiplication of total flux $F$ and polarization degree $P$) in Fig.~\ref{Fig:Center}, to be compared with the same image in total flux (Fig.~\ref{Fig:1995_I}). The reason is that polarized flux maps are known to present better/sharper contrasts to total flux images, where part of the emission is scattered by diffuse material along the line-of-sight or nearby the object of interest. The polarized flux has the advantage of shaving off this environmental flux to better focus on the emitting, scattering and absorbing regions associated to the source itself. 

Thanks to the polarization vectors superimposed onto the images, we can see that the hot spots of polarization (reaching more than 50\% in many pixels) do not correlate with the peak of total flux emission. On contrary, while the maximum of $F$ is observed at the apex of the outflows, high polarization degrees are measured at the extremities of the biconical outflows. However, the degree of polarization is not uniform at the end or within the winds, but seems to be maximum only at the outer edges of the conical winds, leaving the center of the winds relatively little polarized. Such X-shape is a smoking gun evidence for the hollowness of the envelope, according to what was postulated by many authors for, e.g., young stellar objects \citep{Caratti2016}, protostars \citep{Lee2022} or even AGNs \citep{Elvis2000}. 

More importantly, there is a zone at the base of the oppositely directed outflows that is very little polarized in comparison to the outflows. This creates a rectilinear zone about 0.25~arcseconds wide which corresponds entirely to the position of the minimum flux observed in total flux (see Fig.~\ref{Fig:1995_I}, although this zone is only 0.1~arcseconds wide in total flux). An immediate conclusion is that something is obscuring the inner region of LK-H$\alpha$-233. This dark lane, both seen in polarization and total flux, is consistent with the presence of a circumstellar disk or a dust torus, such as postulated by \citet{Melnikov2008}. No polarization vectors are associated with the dark lane, indicating that little-to-no photons directly escape through the dusty circumnuclear obscurer towards us. The optical depth of the dark lane is thus larger than unity (see, e.g., the simulations from \citealt{Murakawa2010} and \citealt{Marin2012}).

Another clear aspect of the polarized flux map is that it reveals a structure that is a bit narrower than in the total flux map, with sharper contours for the bipolar ejecta. LK-H$\alpha$-233 truly resemble to a butterfly in polarized flux : the outflows geometry is precisely revealed, together with the central zone where obscuration takes place. The conical structure of the wind, apparently launched from a compact zone, is striking. It's half-opening angle with respect to the the central axis of the winds is measured to be $\sim$ 44$^\circ$ and $\sim$ 50$^\circ$ for the western (blueshifted) and eastern (redshifted) outflows, respectively. In turns, in the hypothesis that the outflows fill the solid angle left by the obscuring protoplanetary dusty disk, it constrains the disk half-opening angle to about 40 - 46$^\circ$  (with respect to the disk midplane).

\subsection{Locating the central star}
\label{Analysis:center}

We have found that the polarization of LK-H$\alpha$-233 shows a centro-symmetric pattern (see the medium resolution Figs.~\ref{Fig:App_1991b}, \ref{Fig:App_1994b} and \ref{Fig:App_1995b}), something that was already established with single epoch and lower spatial resolution observation, such as with the 0.5~arcsec polarization maps of \citet{Aspin1985}. The overall
distribution of the polarization vectors is, in fact, quite close to the point-source scattering case, something that we can use to determine with precision the position of the emission source, even if it is hidden behind a thick veil of dust and gas. Indeed, when a photon undergoes single scattering from a point source onto an ionized polar outflow, the resulting polarization has an electric vector position angle perpendicular to the direction of the source. Taking into account the full uncertainties in the measured polarization angle of pixels with $\left[\text{S/N}\right]_P \geq 5$, and under the assumption of single scattering onto an ionized medium within a single plane, we can derive the location of the hidden source of emission following the procedure already detailed in \citet{Kishimoto1999} and \citet{Barnouin2024} -- see also \citet{Marin2023} for a similar use of this technique in the case of the Galactic center.

The deduced location of the star in LK-H$\alpha$-233 is shown in Fig.~\ref{Fig:Center}, using the 1995 observations, i.e. the less cropped out of the two post-COSTAR pointings. The location of the hidden star, whose coordinates are indicated on the figure, is precisely at the base of the western outflow, near the location of the brightest flux patches seen in total and polarized fluxes, at the border of the dark lane that is bisecting the Herbig Ae star. The fact that the star is not closer to the dark lane's plane tells us that the inclination of the whole structure is lower than 90$^\circ$ (i.e. perpendicularity).

\subsection{Aperture-dependent polarimetry}
\label{Analysis:aprture}

Now that we have revealed the location of the hidden star, we can simulate a circular aperture onto our 1995's map, centered on the star, and increase its radius to see how the polarization of the system evolves. By expanding the aperture, we aim at determining how different regions of the system contribute to the overall polarization signal and, maybe, identify various scattering regions. Indeed, as the aperture grows, changes in polarization degree and angle might reveal the presence of different scattering structures that affect the light from the hidden star. 

\begin{figure}
\centering
\includegraphics[width=1\columnwidth]{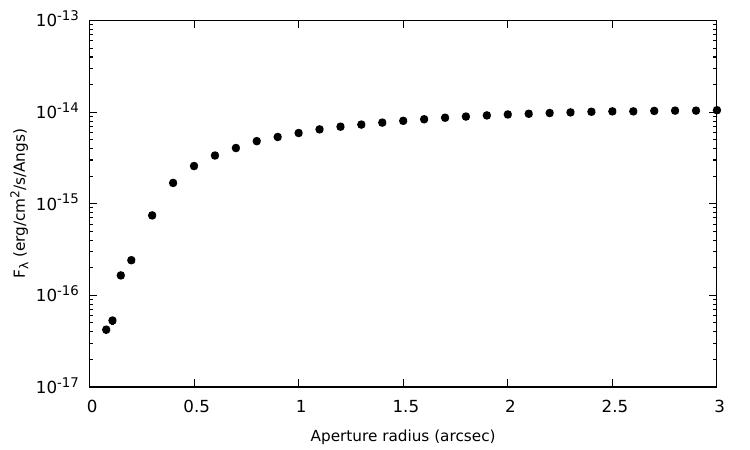}
\includegraphics[width=1\columnwidth]{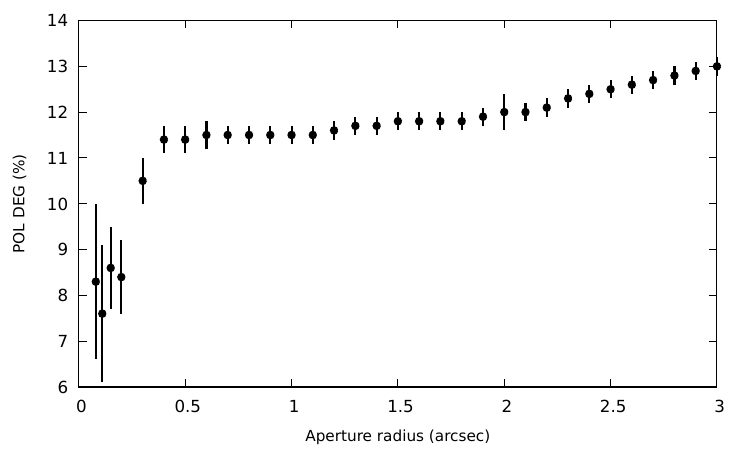}
\includegraphics[width=1\columnwidth]{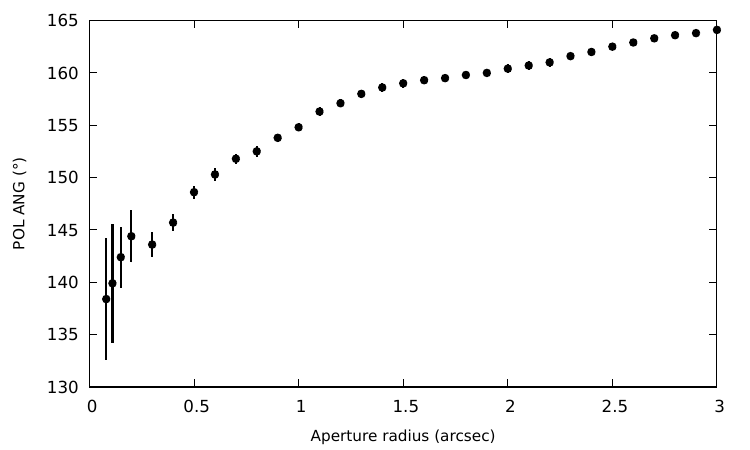}
\caption{Variation of the total flux (top), polarization degree (middle) and polarization angle (bottom) as a function of increasing aperture radii, centered on the location of the hidden star. Error bars are indicated but they are sometimes too small to be visible.}
\label{Fig:Aperture}%
\end{figure}

We show the results of this exercise in Fig.~\ref{Fig:Aperture}. Focusing on the total flux first (top panel), we see that, with increasing aperture radii, the amount of photons we detect steeply rises in the first 0.5~arcsecond around the hidden star. This is logical since the dark line is only 0.25~arcseconds wide. Scattering on the base of the outflows then allows radiation to escape thanks to perpendicular scattering, strongly increasing the observed total flux. For large apertures, the flux seems to plateau at a value of about 1.2 $\times$ 10$^{-14}$ erg cm$^{-2}$ s$^{-1}$ \AA$^{-1}$, which is the value measured for the full field-of-view. More interesting, the polarization degree (Fig.~\ref{Fig:Aperture}, middle panel) shows a modest value, about 8\% in the first 0.25~arcseconds around the hidden star, then jumps to a value larger than 11\% at 0.5~arcseconds. After which, the degree of polarization increases as one moves away from the center of the object, where (single) diffusion is most effective in obtaining a high degree of polarization. This behavior is actually observed in the polarization position angle, which rotates slowly as the aperture radius increases. It starts from about 140$^\circ$ and varies up to about 165$^\circ$ at the largest apertures. When comparing these polarization angles to the outflow position angle (60 – 70$^\circ$ according to \citealt{Corcoran1998}, or 66$^\circ$ $\pm$ 1$^\circ$ as derived in this study), we find that the polarization vectors are approximately perpendicular to the outflow direction across all apertures. However, the fact that there is a median difference of about 16$^\circ$ with respect to orthogonality might indicate that the dust disk might be slightly misaligned with respect to the biconical outflows.

\subsection{Scanning the various polarized regions}
\label{Analysis:scan}

\begin{figure}
\centering
\includegraphics[width=\columnwidth]{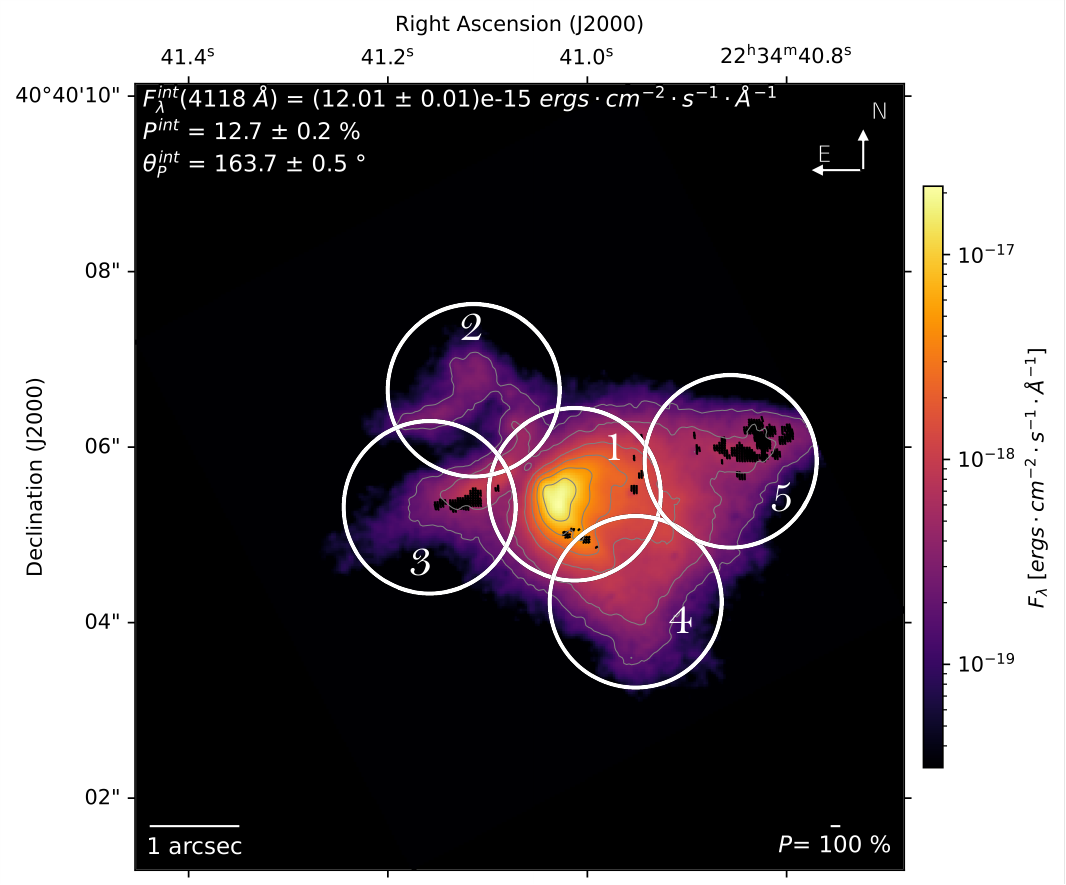}
\caption{The five different zones of LK-H$\alpha$-233 investigated in greater details. The map is the one taken in 1995 (same as Fig.~\ref{Fig:1995_I}). Each white circle as an aperture radius of 1~arcsecond.}
\label{Fig:Frames}%
\end{figure}

\begin{table}
\centering                                      
\begin{tabular}{c c c c}          
\hline\hline                        
\textbf{Region} & \textbf{Total flux} & \textbf{$P$} & \textbf{$\theta$} \\    
~ & \textbf{(ergs~cm$^{-2}$~s$^{-1}$~\AA$^{-1}$)} & \textbf{(\%)} & \textbf{($^\circ$)} \\ 
\hline  
1 & 8.2 $\times$ 10$^{-15}$ $\pm$ 1.2 $\times$ 10$^{-17}$ & 11.4 $\pm$ 0.3 & 153.9 $\pm$ 0.6\\
2 & 2.8 $\times$ 10$^{-16}$ $\pm$ 2.3 $\times$ 10$^{-18}$ & 32.0 $\pm$ 1.3 & 122.8 $\pm$ 1.2\\
3 & 4.7 $\times$ 10$^{-16}$ $\pm$ 2.5 $\times$ 10$^{-18}$ & 52.2 $\pm$ 1.0 & 8.7 $\pm$ 0.5\\
4 & 1.3 $\times$ 10$^{-15}$ $\pm$ 3.0 $\times$ 10$^{-18}$ & 18.3 $\pm$ 0.4 & 130.9 $\pm$ 0.7\\
5 & 1.4 $\times$ 10$^{-15}$ $\pm$ 3.0 $\times$ 10$^{-18}$ & 37.5 $\pm$ 0.4 & 7.0 $\pm$ 0.3\\
\hline                                          
\end{tabular}
\caption{Flux and polarization of the five regions represented by the white circular apertures on Fig.~\ref{Fig:Frames}.} 
\label{Tab:P_zones}     
\end{table}

From the various maps shown in this paper, it is evident that the total and polarized fluxes are spatially dependent in LK-H$\alpha$-233. To explore in greater details its complexity, we decided to divide our 1995's map into five zones, covering most of the source. Each zone is the focus of a synthetic 1~arcsecond radius circular aperture, see Fig.~\ref{Fig:Frames}, labeled 1 (the core), 2 (upper extremity of the eastern outflow), 3 (lower extremity of the eastern outflow), 4 (lower extremity of the western outflow) and 5 (upper extremity of the western outflow). Their integrated flux and polarization are reported in Tab.~\ref{Tab:P_zones}.

Individually, the different regions tell us different stories. Region 1 concentrates most of the source flux ($\sim$ 8.2 $\times$ 10 $^{-15}$ ergs~cm$^{-2}$~s$^{-1}$~\AA$^{-1}$, see Tab.~\ref{Tab:P_zones}), as already discussed in Sect.~\ref{Archives}, and imposes its $P$ and $\theta$ at larger fields-of-view. This is logical as it corresponds to the base of the bipolar winds, where the near-ultraviolet emission from the western component is maximum and where spatially-dependent, 0 - 20\%, polarization is detected, giving rise to high polarized fluxes (see Fig.~\ref{Fig:App_1995b}). The fact that $P$ is "small" ($\sim$ 11\%) at the peak of the total flux may be due to to two distinct, yet compatible, reasons. On one hand, this could be the natural consequence of photons scattering multiple times before escaping, as expected near the highest density regions \citep{Murakawa2010b}. This naturally would explains the increase in polarization as one looks further out in the bipolar nebula as single scattering becomes dominant. On the other hand, it might indicate that the base of the outflow is dynamically active, with many interacting cloudlets or filaments of ionized matter, gas, and dust, giving rise to canceling polarization vectors at spatial scales inferior to the resolution of our HST/FOC maps. Finer spatial resolution polarimetric map would help to distinguish which is the dominant mechanism. 

Region 2, the upper extremity of the eastern outflow, has a flux that is more than an order of magnitude fainter than region 1 (the core). On the contrary, $P$ is greatly higher at this location, were the ejected medium is probably less perturbed than in the core and where single scattering, from an optically thinner medium, would naturally lead to less depolarization \citep{Melnikov2008}. Region 3, the lower extremity of the eastern outflow, also show a dim flux. The integrated polarization degree is surprisingly high, more than 52\%, again indicating an optically-thin medium that is outflowing without strong internal perturbations, reinforcing the validity of the hypothesis used in Sect.~\ref{Analysis:center} to calculate the supposed position of the obscured star. Region 4, the lower extremity of the western outflow, is moderately bright ($\sim$ 1.3 $\times$ 10 $^{-15}$ ergs~cm$^{-2}$~s$^{-1}$~\AA$^{-1}$) and its polarization is slightly higher than in the core region 1. Finally, region 5, the upper extremity of the western outflow, presents not only a moderate brightness ($\sim$ 1.4 $\times$ 10 $^{-15}$ ergs~cm$^{-2}$~s$^{-1}$~\AA$^{-1}$) but also high polarization degrees ($\sim$ 38\%). For those four regions, the observed polarization angle is always perpendicular to the symmetry axis of the system, once their deviation from the central axis of the winds is taken into account.

\subsection{Variability of the polarization}
\label{Analysis:variability}

\begin{figure*}
\centering
\includegraphics[width=1\columnwidth]{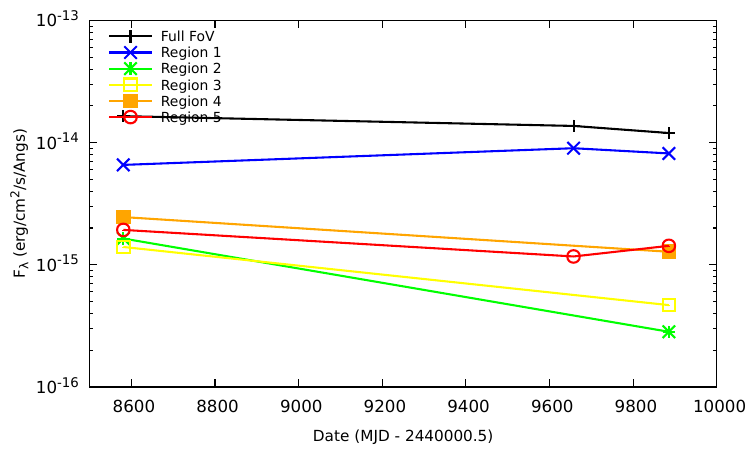}
\includegraphics[width=1\columnwidth]{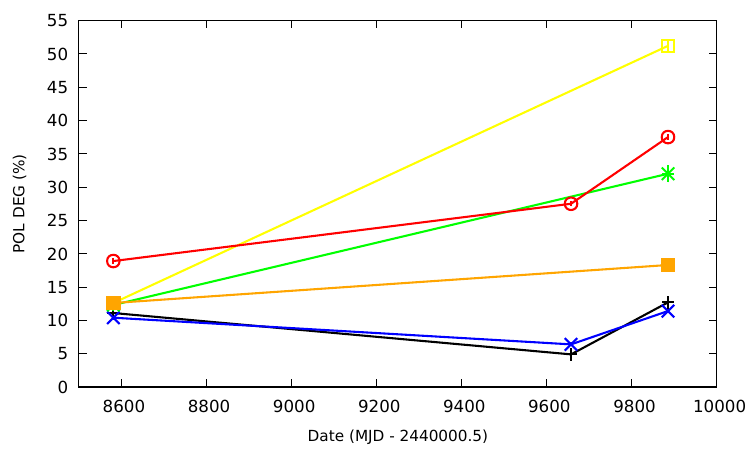}
\includegraphics[width=1\columnwidth]{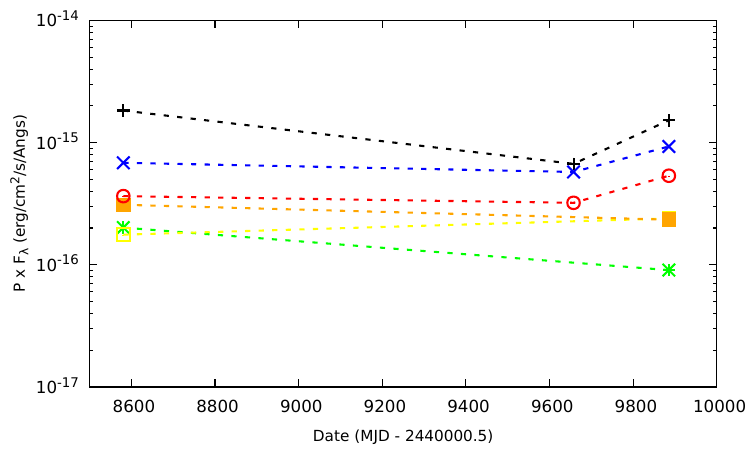}
\includegraphics[width=1\columnwidth]{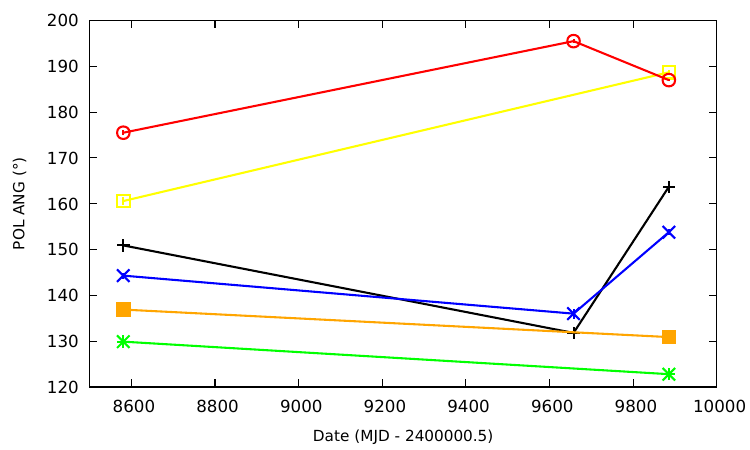}
\caption{Variability of LK-H$\alpha$-233 in total flux (top-left), polarized flux (bottom left), polarization degree (top-right) and polarization angle (bottom right) as a function of time (in modified Julian date). We accounted for the three HST/FOC observations but, due to the pointing error which cut off part of the object in 1994, some points are unavailable in the second observation. Except for the black crosses (full field-of-view, 7 $\times$ 7~arcseconds), all the other data points correspond to 1~arcsecond aperture radius integration windows at the five different regions presented in Fig.~\ref{Fig:Frames}. Error bars are indicated but they are often too small to be visible.}
\label{Fig:Variability}%
\end{figure*}

As previously mentioned in Sect.~\ref{Archives:1994}, Herbig Ae/Be stars are known to vary both in total flux and polarization, with timescales ranging from months to years \citep{Perez1993,Jain1995,Mendigutia2011}. \citet{Jain1995}, in particular, noted that "variations in the polarization position angle are not always correlated with variations in the degree of polarization". \citet{Waters1998} also reported that "variations in polarization are often correlated with the deep photometric minima, and they can be explained in terms of dense dust clouds obscuring the light from the star and allowing only light scattered by dust particles to escape." Now that we have reduced and analyzed both the complete images and different subregions on the three polarimetric dataset, we can examine in greater details the intrinsic variability of the source.

We plot in Fig.~\ref{Fig:Variability} the integrated total flux (top left), polarized flux (bottom left), polarization degree (top right) and polarization angle (bottom right) for the total field-of-view of the HST/FOC observations, but also for the five regions defined and examined in Sect.~\ref{Analysis:scan}. In the case of the full field-of-view, the integration was made on 7 $\times$ 7~arcseconds windows, while for the five subregions, the integration windows were only 1~arcsecond in radius. Due to the pointing error which cut off part of the object in 1994, some points are unavailable in the 1994's observation (regions 2, 3 and 4). 

We find that the variability of LK-H$\alpha$-233 is more complex than expected :
\begin{itemize}
    \item Focusing on the results from the entire object integrated over the whole field-of-view, we see that the total flux marginally decreases with time, while the polarized flux shows a stronger decrease from 1991 to 1994, and then an increase in the last year. This is due to the variations of polarization we observe, with $P$ being the lowest in 1994. Such variation of the polarization degree is associated with a rotation of the polarization position angle. This seems to contradict the findings of \citet{Jain1995}, but one should remember that the 1994 and 1995 observations present cropped images of the object and therefore values that should not be taken at face value. 
    \item Interestingly, the temporal behavior of the central region (region 1) follows almost the same trend as the complete field-of-view, except for the total flux, which shows an increasing trend between 1991 and 1995, followed by a decrease. yet, the polarization degree is minimal in 1994 and associated with a rotation of the polarization angle too. As a consequence, the behavior of the full field-of-view image seems to be constrained by the core region and, despite the Herbig Ae/Be system being cropped out, trustable. The variations in the polarization position angle are correlated with variations in the degree of polarization.
    \item Region 2 (upper extremity of the eastern outflow) has only two points, but the global trend between 1991 and 1995 is a decrease in both total and polarized flux, associated with a steep rise in polarization (from 12.3\% $\pm$ 0.4\% to 32.0\% $\pm$ 1.3\%) and a $\sim$ 7$^\circ$ rotation of the polarization angle.
    \item Region 3 (lower extremity of the eastern outflow) has also two points but their temporal behavior differs from region 2 : the total flux decreases with time but the polarized flux increases. In fact, the rise in $P$ is even stronger than for region 2 (from 12.6\% $\pm$ 0.4\% to 51.2\% $\pm$ 1.0\%) and the rotation of the polarization angle is about 28$^\circ$.
    \item Region 4 (lower extremity of the western outflow), two points only too, behaves exactly like region 2, with similar changes in polarization degree and angle, although the gain in $P$ is slightly bigger (6 percentage points between the two observations).
    \item Finally, region 5 (upper extremity of the western outflow), has three points and follows its own path : its total flux first decreases between 1991 and 1995 but then increases in 1995. The polarized flux of region 5 follows the same trend, constrained by a continuous rise in $P$ over three years. However, in contrast to all the other regions investigated and for which we have enough data point, its polarization position angle rotates but with a different sign with respect to the full field of view and region 1 : first the polarization angles gains +20$^\circ$ between 1991 and 1995, but then it changes by about -9$^\circ$.
\end{itemize}

Our multi-epoch polarimetric observations of LK-H$\alpha$-233 reveal a complex spatial and temporal variability in both total and polarized flux. While the global trends are largely governed by the central region, the outflows exhibit distinct polarization behavior, suggesting asymmetries in dust distribution and scattering processes. The observed changes in polarization degree and position angle indicate evolving circumstellar structures, possibly linked to variable dust obscuration or disk reconfiguration \citep{Waters1998,Marin2017}. Observing Herbig Ae/Be stars at high spatial resolution is thus a necessity to disentangle and quantify the contribution of the central object from the large-scale outflows.

\section{Discussion}
\label{Discussion}

\subsection{The inclination of the system}
\label{Discussion:inclination}
Thanks to the data collected during this analysis, we have constrained the geometry of both the accretion disk (flared, half-opening angle of 40 - 46$^\circ$ with respect to the disk midplane) and of the outflows (biconical, hollow, collimated by the circumstellar dust, half-opening angle of 44 - 50$^\circ$ with respect to the the central axis of the winds). The outflows are scattering the obscured starlight towards us and we can use this information to estimate, at the first order only, the inclination of the whole system.  

In the idealized case where scattering occurs on free electrons, and no significant unpolarized (diluting) component is present, the linear polarization degree $P$ can be approximated by the classical equation for Thomson scattering: $P$ = (1-$\mu^2$)/(1+$\mu^2$), where $\mu = \cos{\theta}$ and $\theta$ is the scattering angle \citep{Kishimoto1999}. This expression derives from the ratio of the Stokes parameters: the numerator corresponds to the polarized intensity (Stokes $Q$ and $U$), and the denominator to the total intensity (Stokes $I$). It assumes that the scattered light originates from a compact illuminating source, such that the scattering volume subtended by each image pixel occupies a small solid angle as seen from the source. It also assumes single scattering and a low optical depth medium. Clearly, the core region of the system is dominated by multiple scattering in dense environment, so we cannot rely on polarimetric measurement from this spatial location. However, we can use the polarization measured at the extremities of the winds as a proxy to calculate the inclination of the object, since single scattering seems to dominate at those location (see Sect.~\ref{Analysis:aprture}).

On our polarization maps from all three periods, we measure a polarization degree that ranges from 60 to 80\% at the ends of the winds, in a 0.5~arcsecond circular aperture. This translates into an inclination of 60 -- 70$^\circ$. This is a rough estimate but it is in agreement with the values obtained by \citet{Perrin2009}, who estimated that the disk is inclined by 65$^\circ$ $\pm$ 5$^\circ$ relative to the plane of the sky using numerical radiative transfer disk models. On the other hand, \citet{Bastien1990}, also using radiative transfer models, rather determined an inclination of 85 - 90$^\circ$. Because we see the eastern, opposite outflow (redshifted, see \citealt{Melnikov2008}) and because we don't see the central pre-main sequence star, the system must have an inclination larger than the disk half-opening angle, so that our line-of-sight is intercepted by dust. It means that even if we cannot be sure about the global inclination of the system, we can still derive a pretty strict upper limit to the inclination of the source, that must have an orientation below the half-opening angle of the disk, i.e., $\ge$ 45$^\circ$ relative to the plane of the sky. 

To highlight the significance of determining the system's orientation, we can estimate what fraction of the stellar flux is represented by the measured fluxes — resulting from scattering in the polar winds — using the inclination constraints we derived. Based on the numerical work of \citet{Marin2012}, where a central, isotropic, compact illuminating source of optical radiation is surrounded by an optically-thick, dusty disk that only allows an electron-filled wind to escape along the polar (unobstructed) direction, a geometry that mimics very well the case of Herbig Ae/be star systems, we can provide an educated guess. Looking at their Fig.~19 (bottom), an inclination larger than 45$^\circ$ means that only 1 to 10\% of the original flux is scattered in the direction of the observer, depending on the exact inclination of the system. This implies that the intrinsic flux of the star inside LK-H$\alpha$-233 is at least a factor 10, if not a factor 100 greater than 10$^{-14}$ ergs~cm$^{-2}$~s$^{-1}$~\AA$^{-1}$. This estimate assumes that the dominant polarigenic opacity in the wind is electron scattering. However, if dust scattering dominates (as it seems to be the case for this object, see \citealt{Vrba1979}, \citealt{Aspin1985} and \citealt{Leinert1993}), \citet{Goosmann2007} have shown that the observed flux under such conditions may represent only $\sim$ 0.1\% of the intrinsic emission. In that case, the true stellar flux could be up to 1000 times higher than observed, implying an intrinsic flux closer to a few 10$^{-11}$ ergs~cm$^{-2}$~s$^{-1}$~\AA$^{-1}$ at 4118~\AA.

\subsection{How are the winds created?}
\label{Discussion:Winds}

We have found that the winds of LK-H$\alpha$-233 are hollow, brighter in total flux at the wind base but stronger in polarization at the wind maximal extension. The geometry of the ejecta, revealed in polarized flux, has an X-shape. Its half-opening angle is 44 - 50$^\circ$ with respect to the the central axis of the winds. Scattering prevails in the winds, as shown by the centro-symmetric pattern of the polarization vectors. These particularities can potentially help us to better constrain what could be the ejection mechanism at work here.

Although the specific mechanisms of wind launching and collimation differ, disk-born winds \citep[e.g.][]{Ustyugova1995}, X-winds originating near the disk–magnetosphere boundary \citep[e.g.][]{Romanova2009} and stellar winds from open magnetic field lines anchored to the star’s surface \citep[e.g.][]{Decampli1981} can produce hollow winds, but their range of half-opening angles differ. Disk winds are launched from a range of radii on the accretion disk, resulting in a wide-angle outflow. The exact angle can depend on the strength and configuration of the magnetic field, the rate of accretion and the specific disk properties, but the typical range of values goes from 6$^\circ$ to 60$^\circ$ \citep{Blandford1982,Pudritz2007}, with most of the mass flux occurring in a conical shell with half-opening angle of 25$^\circ$ – 30$^\circ$ \citep{Villiers2005}. In the case of X-winds, because they originate near the disk-magnetosphere boundary, they are thought to be more tightly collimated compared to disk-born winds. The strong magnetic fields in this region help focus the wind into a narrower cone of half-opening angles typically ranging from 10$^\circ$ to 30$^\circ$ \citep{Shu1994,Lovelace2014}. Finally, stellar winds launched from the magnetic poles of the star can be highly collimated, especially if the magnetic field lines are well-ordered and strong. However, if the magnetic field is weaker or more chaotic, the winds can have a broader opening angle, leading to a typical range of half-opening angles of 10$^\circ$ to 40$^\circ$ \citep{Shore1987}. With a measured half-opening angle measured between 44$^\circ$ and 50$^\circ$, our analysis would rather lean towards disk-born winds, but it should be noted that it is very likely that each model can reproduce these values with the right set of (sometimes extreme) parameters. Therefore, the opening angle alone does not provide definitive evidence for identifying the wind launching mechanism.

If we investigate the polarization resulting from each models, then we may get a more convincing proof. As shown by \citet{Murakawa2010b}, disk-born winds exhibits strong total flux signals at the base of the wind and high linear polarization degrees inside the wind, with higher degrees found at the wind maximal extension (up to 60\%). The resulting polarization pattern is centro-symmetric around the central illuminating source, consistent with scattering in a hollow, axisymmetric geometry. This pattern extends across the region defined by the detected polarized flux and does not necessarily encompass the full extent of the wind as traced by other tracers (e.g., forbidden lines or molecular outflows), but it does reveal the geometry of the arcsecond-scale component of the outflow. In models where magnetic fields dominate (e.g., X-winds or stellar winds), the observed polarization can depend critically on the alignment of dust grains with magnetic field lines. In the presence of a strong and ordered magnetic field, non-spherical dust grains can align with the magnetic field via radiative alignment torque mechanisms \citep[e.g.][]{Lazarian2007,Hoang2008}, producing a measurable linear polarization signal through dichroic extinction or emission. In contrast, in turbulent or weak magnetic fields, grain alignment may be suppressed or randomized, leading to significantly reduced net polarization. Therefore, high degrees of linear polarization can be a signature of well-ordered magnetic structures (that should be accompanied with detectable synchrotron-induced circular polarization \citep{Hubrig2004}), while low polarization can indicate disordered or tangled fields. Unfortunately, no theoretical prediction about the expected degree of linear polarization exists for the X-winds nor the stellar winds from open magnetic field lines anchored to the star’s surface models, so it is not possible to determine if they match our data. Yet, to reproduce the observed polarization properties of LK-H$\alpha$-233, it would require randomly oriented magnetic fields at the apex of the winds (near the central star) and highly ordered ones at the end of the ejecta, which is rather counter-intuitive : field lines tend to break up and become more disordered as they move away from their point of origin, similar to what is observed in AGN jets, where relativistic collimated jets fragment and form individual blobs as they moves away from the vicinity of the supermassive black hole where they were generated. \citep{Begelman1984,Pino2005}.

Overall, it is not possible to determine with certainty what is the mechanism responsible for the creation of the winds observed for the subject of this study, but the few leads that we have tend to favor the creation/ejection of winds from the surface of the dust disk that belts the star. More theoretical work on the different models is needed to determine their polarization properties before concluding definitively.

\subsection{An intriguing resemblance to AGNs}
\label{Discussion:AGN}

Herbig Ae/Be systems typically consist of several key components that play crucial roles in the star formation and early stellar evolution processes. These components include the young, pre-main-sequence star itself, the circumstellar disk and bipolar outflows, sometimes associated with jets. Seen from the top, the star is unobscured and its characteristics can be directly measured. Seen from the equatorial plane, the star is hidden behind the circumstellar disk and only a fraction of the radiation emitted by the star is observed, scattered onto the outflows. In the former case, the observed polarization is often small, often parallel to the outflows or jets, while in the former case, the polarization can reaches several tens of percent, with a polarization position angle perpendicular to the ejecta \citep{Bastien1987,Vink2002,Maheswar2002}. Scattering is the main mechanism to explain the observed polarization in the ultraviolet and optical, and dichroic emission and absorption take precedence in the infrared.

It is interesting to note that, despite the dramatic differences in scales and power, the case of AGNs is very similar in terms of geometry. The central engine (a supermassive black hole and its accretion disk) is nested inside a circumnuclear, obscuring material that collimates the outflows in the polar directions. In some cases, relativistic, beamed jet are detected \citep{Antonucci1993,Urry1995}. Of course, the circumstellar and circumnuclear disks do not share the same physics and do not contribute to the system in a similar manner, but their geometry and their ability to obscure the source of emission, as well as their effect of collimating ejection winds in the polar directions, contributes to create, at zeroth order, an uncanny resemblance. They also share the same polarization properties, depending on the orientation of the observer \citep{Antonucci1993,Marin2014}. Hollow winds are increasingly detected \citep{Veilleux2005,Congiu2017,Speranza2024} and the flared nature of the disk is also preferred over flat models \citep{Manske1998,Landt2023}. Finally, the current constraints on the half-opening angle to the optically thick, circumnuclear obscurer, obtained thanks to X-ray polarimetry (45 - 55$^\circ$, \citealt{Ursini2023,Marin2024}) are aligned with the value found in this paper (40 - 46$^\circ$). 

AGNs are extragalactic, distant objects, often dim, so imaging polarimetry, even at the exquisite spatial resolution of the FOC, cannot resolve structures smaller than than 50 - 100~pc in the nearest objects, and Nyquist sampling is not possible as the signal to noise in polarization is low despite long observational times \citep{Barnouin2023}. The dusty obscurer, in AGNs, has an outer radius which depends on the wavelength at which we probe the object, but the best constraints in infrared, optical and ultraviolet give a size of approximately a few parsecs \citep{Gonzalez2019}. This is hardly spatially resolvable in nearby, powerful AGNs and well below what we can reach for more distant objects, not even accounting for polarimetry. It is a pity because only polarimetry can tell us about the true geometry of the winds, as well as the forces that collimate them. Thereby, hints to decipher how circumnuclear material can collimate winds, how disks can produce biconical hollow outflows and what is the geometry of the obscurer (flared, clumpy...) might lie in the observation of nearby Herbig Ae/Be stars with high angular resolution imaging polarimetry.

\section{Conclusion}
\label{Conclusion}

We retrieved forgotten high spatial resolution near-ultraviolet polarimetric observations of the Herbig Ae/Be star LK-H$\alpha$-233 in the archival data from the Hubble Space Telescope's Faint Object Camera (HST/FOC). Reducing the data using state-of-the-art pipelines, the observation provided detailed insights into the circumstellar environment, revealing the structure of the star's outflows and dusty disk. We list below the highlights of our analysis:

\begin{itemize}
    \item We obtained total flux and polarization maps at a spatial resolution of 0.0287 $\times$ 0.0287~arcseconds$^2$, significantly higher than previous studies and resolved detailed structures within the circumstellar environment of LK-H$\alpha$-233. 
    \item Medium resolution images (pixel size of 0.1~arcsec) were also produced to better highlight the polarization pattern of the observed radiation.
    \item A dark lane, 0.1~arcsecond wide in total flux, 0.25~arcsecond wide in polarized flux, clearly separating the eastern and western outflows, suggests the presence of a circumstellar disk or dust torus that is both obscuring the pre-main sequence star and collimating the outflows.
    \item High degrees of polarization (up to 80\%) are observed at the extremities of the biconical outflows, whose true morphology is revealed thanks to polarized flux maps. The outflows have a clear X-shape (or butterfly structure), indicating hollow outflows, with maximum polarization at the outer edges.
    \item A clear centro-symmetric pattern is revealed in the outflows by the polarization angles, indicating that scattering is the main mechanism by which photons from the star are reaching the observer. The outflows are thus acting like mirrors, providing a periscopic view of the inner regions of the disk plus star system.
    \item The total flux and polarization present significant variability over the three epochs (1991, 1994, 1995). Variations in polarization degree are always associated with a rotation of the polarization position angle.
    \item We measured the outflows half-opening angles and found $\sim$ 44$^\circ$ and $\sim$ 50$^\circ$ for the western and eastern winds, respectively. In turns, it constrains the dusty disk half-opening angle to 40 - 45$^\circ$ (with respect to the disk midplane). 
    \item At first order, the inclination angle of the observer is found to be 60 -- 70$^\circ$ but a more conservative approach gives a strict upper limit of $\ge$ 45$^\circ$ relative to the plane of the sky to explain all the observed properties of the system.
    \item Finally, from Monte Carlo simulations \citep{Goosmann2007,Marin2012}, the obscured star's intrinsic flux is estimated to be about a few 10$^{-11}$ ergs~cm$^{-2}$~s$^{-1}$~\AA$^{-1}$ at 4118~\AA. 
\end{itemize}

In summary, this work has demonstrated the importance of high spatial resolution polarimetry in studying the circumstellar environments of Herbig Ae/Be stars. However, it would be possible to take the analysis further by having access to similar or better measurements in other wavelengths in order to be able to determine the composition of the winds according to their distance from the central source. Polarimetric images with high angular resolution (comparable or better than that of the FOC) in the optical or infrared bands could determine whether the diffusion observed in the winds is mainly due to electrons (degree of polarization constant with wavelength) or to dust (degree of polarization variable with wavelength). Thus, it would be possible to characterize the dust, both in the winds but also in the disk, whose radius seems to be at the limit of the spatial resolution and signal-to-noise ratio obtained by the HST/FOC. A polarimetric instrument with very high angular resolution (1 $\times$ 1~milliarcsecond$^2$) would undoubtedly make it possible to clearly image and characterize the dust disk of LK-H$\alpha$-233.


\begin{acknowledgements}
The author would like to thank the referees for their insightful comments and suggestions, which significantly improved the quality of this article. The author also extends sincere thanks to Thibault Barnouin, Bruno Catala, and Evelyne Alecian for their valuable feedback, which further contributed to enhancing the manuscript. The author would finally like to acknowledge the support of the CNRS, the University of Strasbourg, the AT-PEM and ATCG. This work was supported by the "Action Thématique Phénomènes Extrêmes et Multi-messagers " (AT-PEM) and the "Action Thématique de Cosmologie et Galaxies" (ATCG) of CNRS/INSU co-funded by CNRS/IN2P3, CNRS/INP, CEA and CNES. 
\end{acknowledgements}

\bibliographystyle{aa} 
\bibliography{biblio}

\begin{appendix}
\onecolumn
\section{Full scale images}
We show here the full scale images of LK-H$\alpha$-233 to better detect the polarization vectors and the pattern they create. 

\begin{figure*}[h!]
\centering
\includegraphics[width=\textwidth]{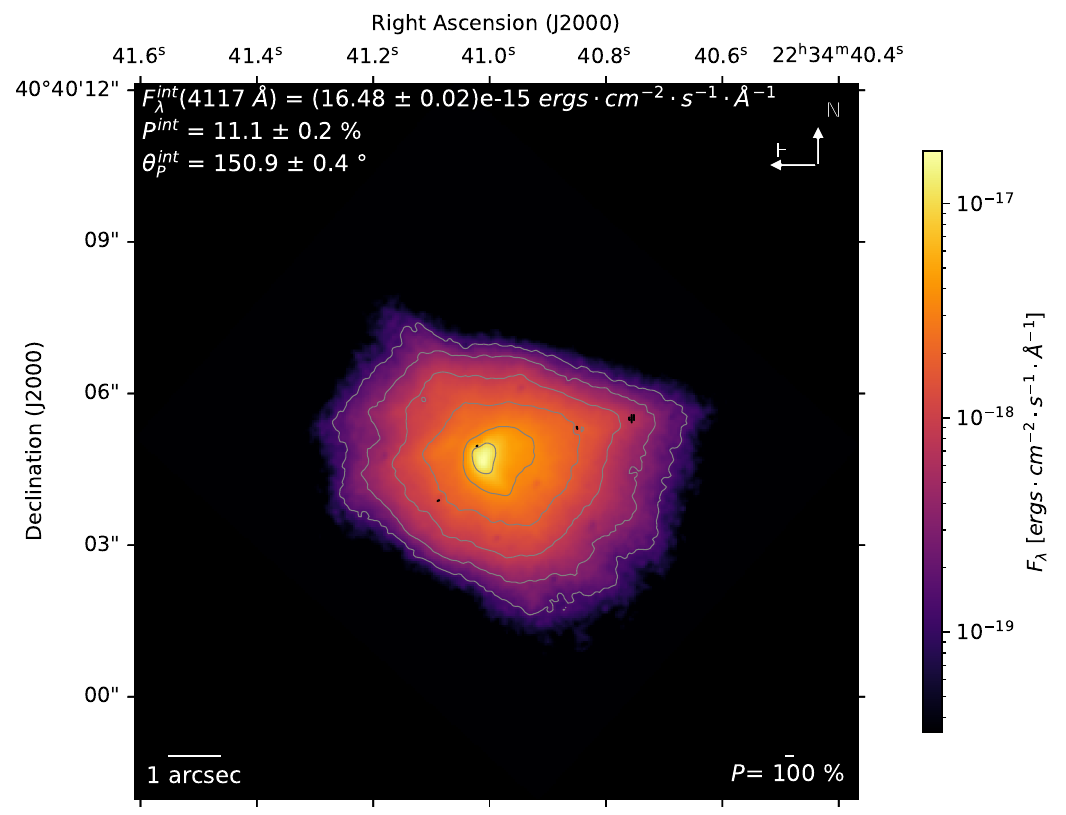}
\caption{1991's pre-COSTAR HST/FOC observation of LK-H$\alpha$-233 resampled according to the Nyquist–Shannon sampling theorem, i.e. 2 $\times$ 2 pixels$^2$ (0.0287 $\times$ 0.0287~arcseconds$^2$) and presented in Fig.~\ref{Fig:1991_I} in a more compressed version.}
\label{Fig:App_1991a}%
\end{figure*}

\begin{figure*}[h!]
\centering
\includegraphics[width=\textwidth]{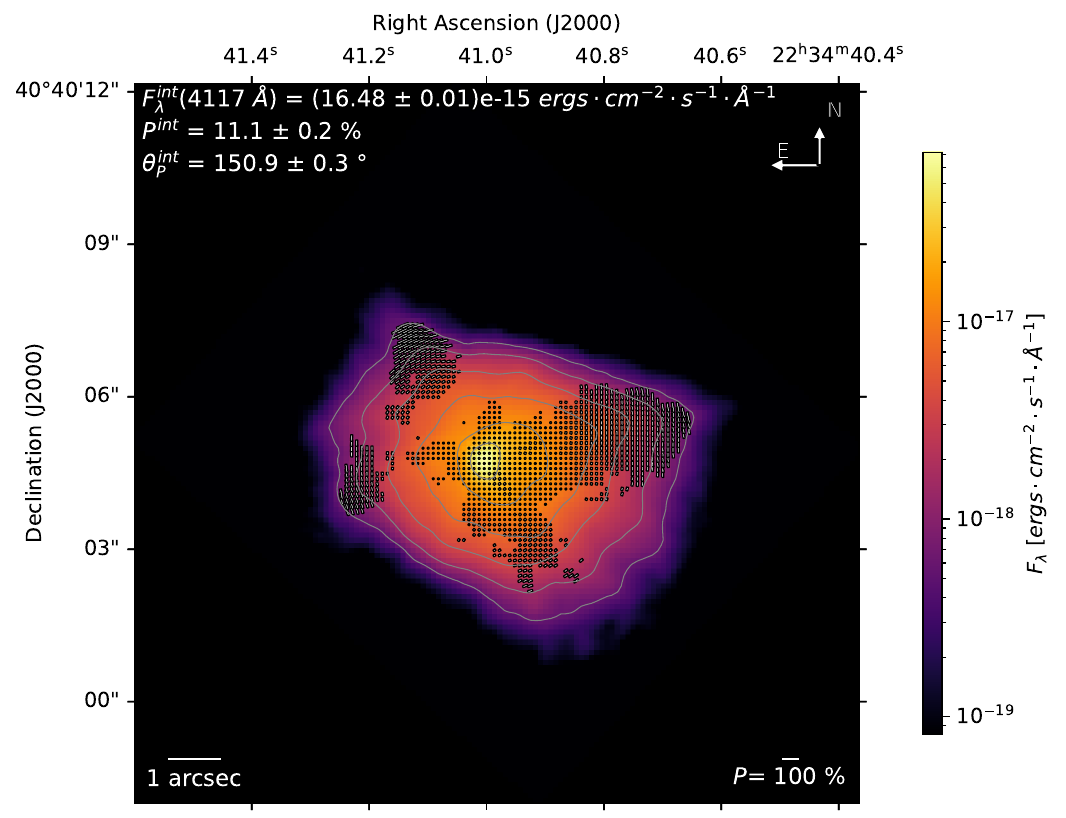}
\caption{Same as Fig.~\ref{Fig:App_1991a} but with a spatial binning of 0.1 arcsecond per pixel.}
\label{Fig:App_1991b}%
\end{figure*}

\begin{figure*}[h!]
\centering
\includegraphics[width=\textwidth]{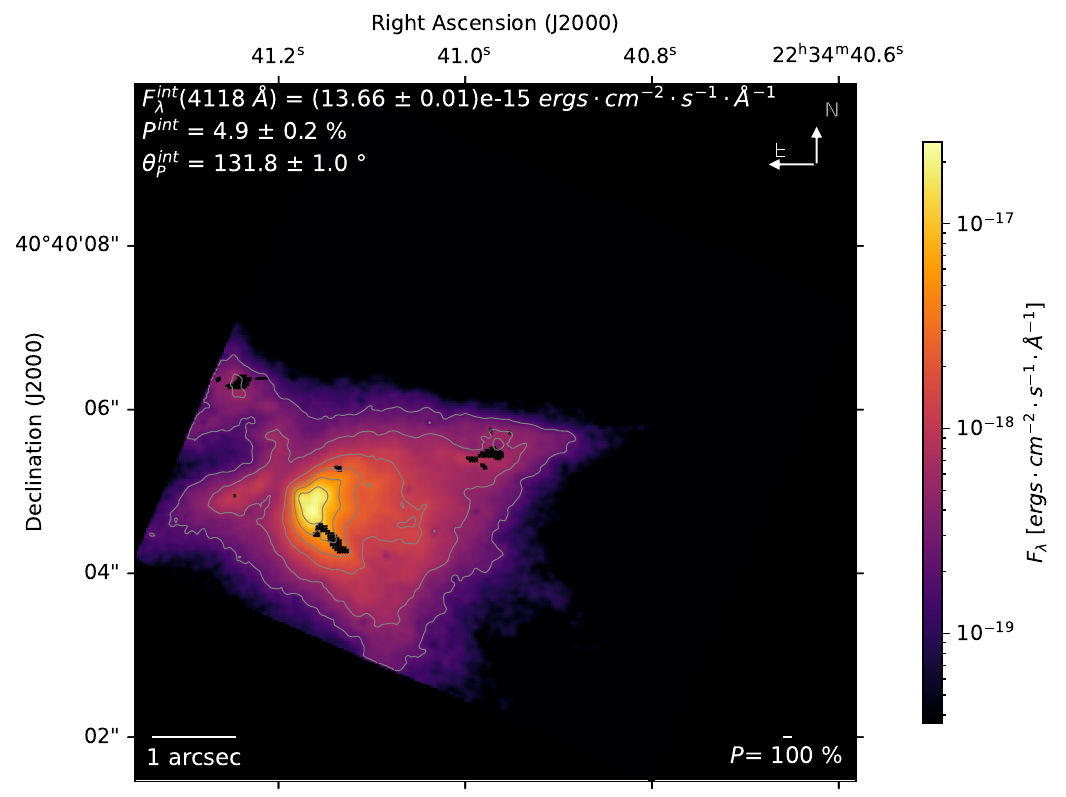}
\caption{1994’s post-COSTAR HST/FOC observation of LK-H$\alpha$-233 resampled according to the Nyquist–Shannon sampling theorem, i.e. 2 $\times$ 2 pixels$^2$ (0.0287 $\times$ 0.0287~arcseconds$^2$) and presented in Fig.~\ref{Fig:1994_I} in a more compressed version.}
\label{Fig:App_1994a}%
\end{figure*}

\begin{figure*}[h!]
\centering
\includegraphics[width=\textwidth]{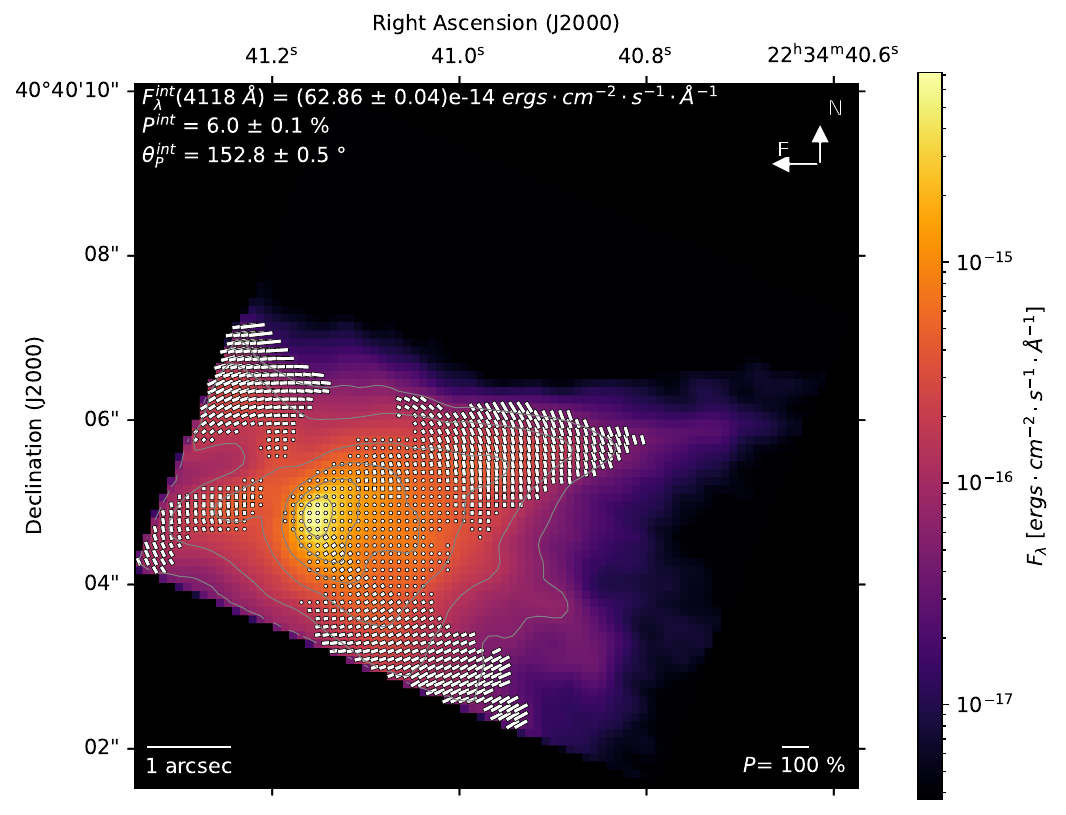}
\caption{Same as Fig.~\ref{Fig:App_1994a} but with a spatial binning of 0.1 arcsecond per pixel.}
\label{Fig:App_1994b}%
\end{figure*}

\begin{figure*}[h!]
\centering
\includegraphics[width=\textwidth]{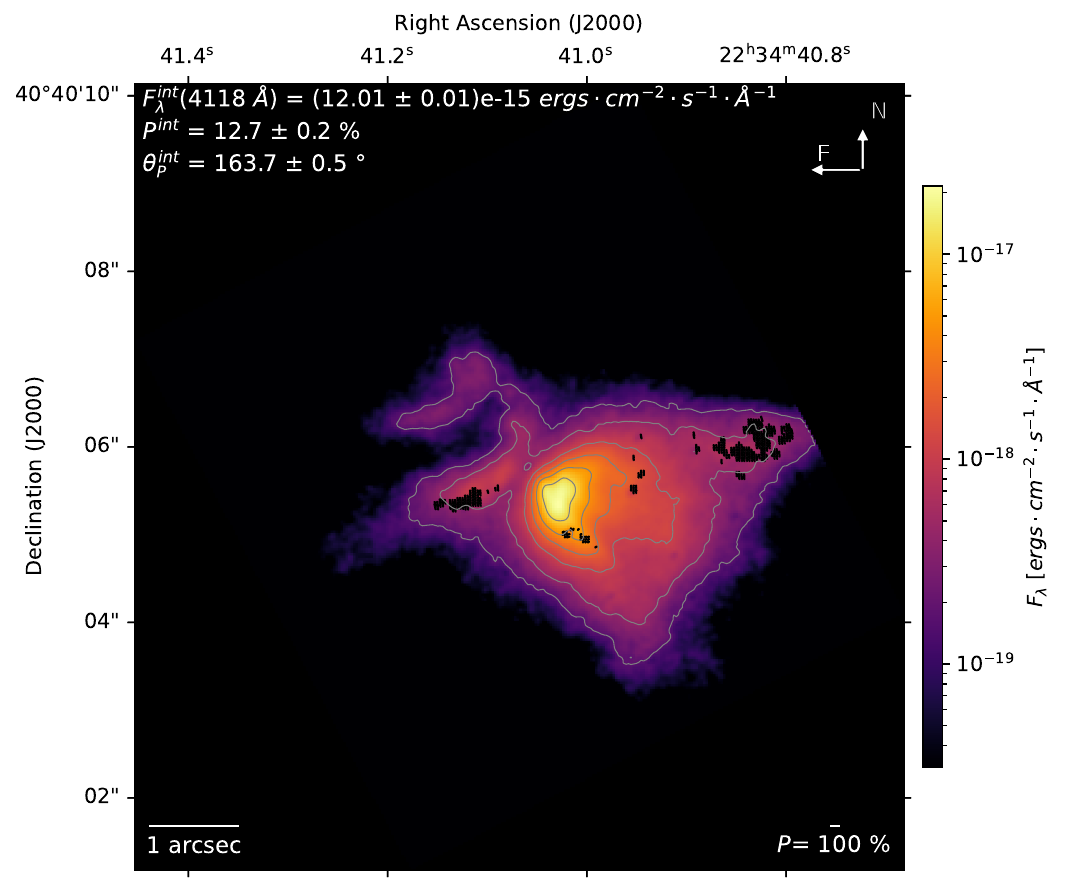}
\caption{1995’s post-COSTAR HST/FOC observation of LK-H$\alpha$-233 resampled according to the Nyquist–Shannon sampling theorem, i.e. 2 $\times$ 2 pixels$^2$ (0.0287 $\times$ 0.0287~arcseconds$^2$) and presented in Fig.~\ref{Fig:1995_I} in a more compressed version.}
\label{Fig:App_1995a}%
\end{figure*}

\begin{figure*}[h!]
\centering
\includegraphics[width=\textwidth]{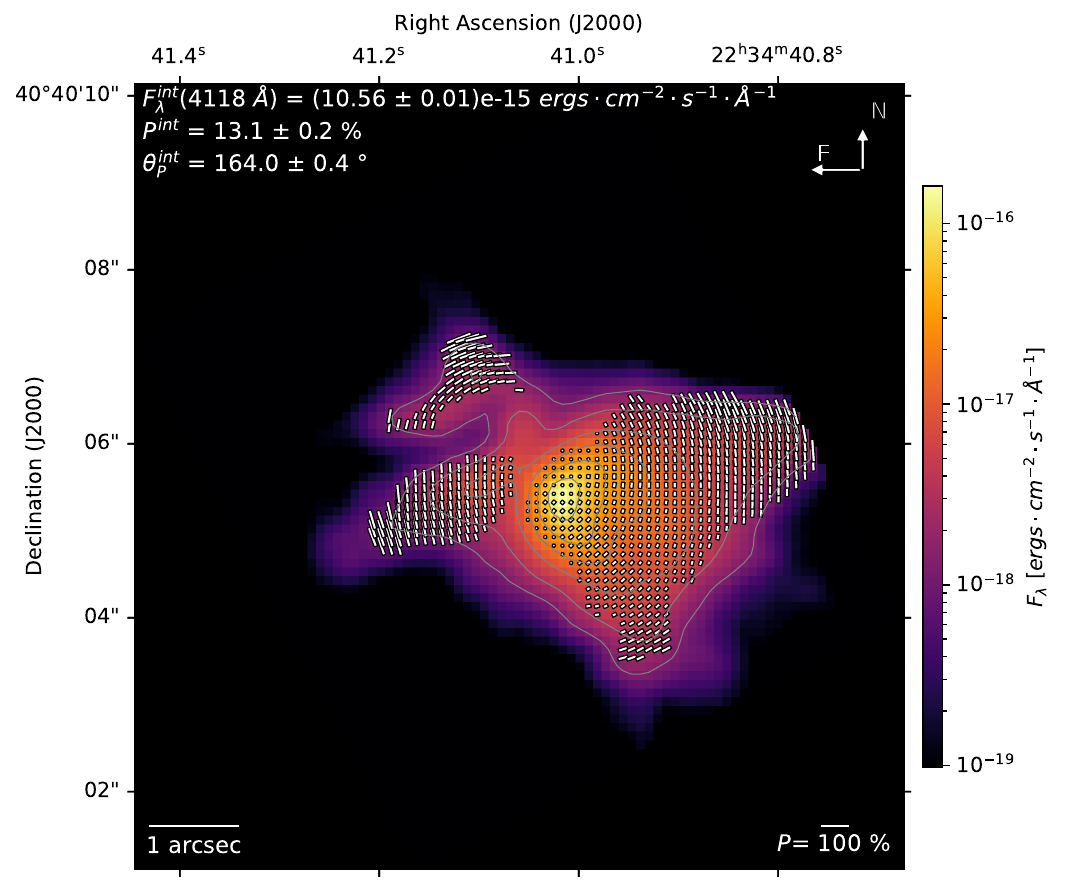}
\caption{Same as Fig.~\ref{Fig:App_1995a} but with a spatial binning of 0.1 arcsecond per pixel.}
\label{Fig:App_1995b}%
\end{figure*}

\end{appendix}

\end{document}